\documentclass{PoS}

\usepackage{lineno}

\title{Rare B Decays Potential of SuperB}

\ShortTitle{Rare B decays potential of SuperB}

\author{Alejandro P\'erez \\
        Laboratoire de l'Acc\'el\'erateur Lin\'eaire (LAL)\\
        E-mail: \email{perez@lal.in2p3.fr}}

\abstract{The study of rare B-decays at SuperB provides unique opportunities to understand the 
Standard Model (SM) and constrain new physics (NP). It is discussed the new physics potential of the 
$B\rightarrow K\nu\bar{\nu}$ and $B\rightarrow K^{*}\nu\bar{\nu}$ system from the proposed 
SuperB experiment with $75{\rm ab^{-1}}$ of data (5 nominal years of data taking).}

\FullConference{35th International Conference of High Energy Physics - ICHEP2010,\\
		July 22-28, 2010\\
		Paris France}

\begin{document}

\section{Introduction}

Rare B decays with a $\nu\bar{\nu}$ pair in the final state are interesting probes of NP, 
since they allow one to transparently study contributions to $Z$ and electroweak penguins.
Furthermore, since the neutrinos escape the detector unmeasured, the $B \rightarrow K^{(*)} + E_{\rm miss}$ 
channel can also contain contributions from other light SM-singlet particles substituting the 
neutrinos in the decay. The $b \rightarrow s\nu\bar{\nu}$ decays are particularly interesting as it is 
possible to formulate model-idenpendent phenomenological analysis. Out of the 
$B\rightarrow K\nu\bar{\nu}$ and $B\rightarrow K^*\nu\bar{\nu}$ decays, there are three observables accessible: 
the corresponding branching ratios and the $K^*$ longitudinal polarization fraction $\langle F_L\rangle$ from 
$B \rightarrow K^* \nu\bar{\nu}$ decays. These three observables only depend on two combinations of the Wilson 
coefficients $C^{\nu}_L$ and $C^{\nu}_R$~\cite{Buras} through the variables 
$\epsilon = \frac{\sqrt{|C^{\nu}_{L}|^2+|C^{\nu}_{R}|^2}}{|(C^{\nu}_{L})^{SM}|}$ 
and $\eta = \frac{-{\rm Re(C^{\nu}_{L}C^{\nu*}_{R})}}{|C^{\nu}_{L}|^2+|C^{\nu}_{R}|^2}$, 
($\eta$ lies in the range $[-\frac{1}{2},\frac{1}{2}]$). The discussed observables can 
be expressed in terms of $\epsilon$ and $\eta$ as follows,
\begin{eqnarray}
\label{eq:Observables}
{\rm Br}(B\rightarrow K^*\nu\bar{\nu})  &=& 6.8\times10^{-6}(1+1.31\eta)\epsilon^2~, \\
{\rm Br}(B\rightarrow K \nu\bar{\nu})   &=& 4.5\times10^{-6}(1-2\eta)\epsilon^2~, \\
\langle{\rm F}_L(B\rightarrow K^*\nu\bar{\nu})\rangle &=& 0.54\frac{(1+2\eta)}{(1+1.31\eta)}~.
\end{eqnarray} 
Table~\ref{tab:Current_results} shows the SM predictions and current experimental upper bounds on branching ratios (${\rm Br}$). 
The experimental bounds on the ${\rm Br}$ can then be translated in excluded areas on the $\epsilon-\eta$ 
plane (green area on the rightmost plot of figure~\ref{fig:Ext_BFsKnunu_Kstnunu}), where the SM corresponds to 
$(\epsilon,\eta) = (1,0)$.

\begin{table}[hbt!]
\begin{center}
\begin{TableSize}
\begin{tabular}{lcc}
\hline
Observable         & SM prediction & Experiment \\
\hline
${\rm Br}(B\rightarrow K\nu\bar{\nu})$                   & $(6.8^{+1.0}_{-1.1})\times 10^{-6}$~\cite{Buras} & $< 80\times 10^{-6}$~\cite{BaBar_meas} \\
${\rm Br}(B\rightarrow K^*\nu\bar{\nu})$                 & $(4.5 \pm 0.7)\times 10^{-6}$~\cite{Buras}       & $< 14\times 10^{-6}$~\cite{Belle_meas} \\
$\langle{\rm F}_L(B\rightarrow K^*\nu\bar{\nu})\rangle$  & $0.54 \pm 0.01$~\cite{Buras}                     & --- \\
\hline
\end{tabular}
\end{TableSize}
\end{center}
\caption{\em SM predictions and experimental $90\%~{\rm C.L.}$ upper bounds for the three $B \rightarrow K^{(*)} \nu\bar{\nu}$ observables.
\label{tab:Current_results}}
\end{table}

\section{The experimental technique and Expectations for SuperB}

The recoil technique has been developed 
to search for rare B decays with undetected particles, like neutrinos, in the final states. 
Its consists on the reconstruction of one of the two B mesons ($B_{\rm tag}$), produced through the 
$e^+e^- \rightarrow \Upsilon(4S) \rightarrow B\bar{B}$ resonance, in a high purity 
hadronic or semi-leptonic final states, allowing to build a pure sample of $B\bar{B}$ events. 
Having identified the $B_{\rm tag}$, everything in the rest of the event (ROE) belongs by default to the signal 
B candidate ($B_{\rm sig}$), and so this technique provides a clean environment to search for rare 
decays. In this analysis, the $B_{\rm tag}$ is reconstructed in the hadronic modes $B\rightarrow D^{(*)}X$, 
where $X = n\pi + mK + pK^0_S + q\pi^0$ ($n+m+p+q < 6$), or semi-leptonic modes $B\rightarrow D^{(*)}\ell\nu$, 
($\ell = e,~\mu$). In the search for $B\rightarrow K\nu\bar{\nu}$ ($B\rightarrow K^*\nu\bar{\nu}$), 
the signal is given by a single track identified as a kaon (a $K^*$ reconstructed in the 
$K^{*0}\rightarrow K^+\pi^-$, $K^{*+}\rightarrow K^+\pi^0/K^0_S\pi^+$ modes) in the ROE.

Even though the expected SuperB~\cite{superB} increase in the instantaneous luminosity of a factor of 
$~100$ already promises significant improvements on the before mentionned rare decays, additional 
activities for detector optimization are currently ongoing. The baseline SuperB detector configuration 
is very similar to BaBar but the boost ($\beta\gamma$) is reduced from 0.56 to 0.28. 
This boost reduction increases the geometrical acceptance and so the 
reconstruction efficiency. Additionally, it is considered the inclusion of a highly performant particle 
identification device (Fwd-PID) based on time-of-flight measurements in the forward region.

\begin{figure}[h!]
\begin{center}
\includegraphics[width=4.9cm,keepaspectratio]{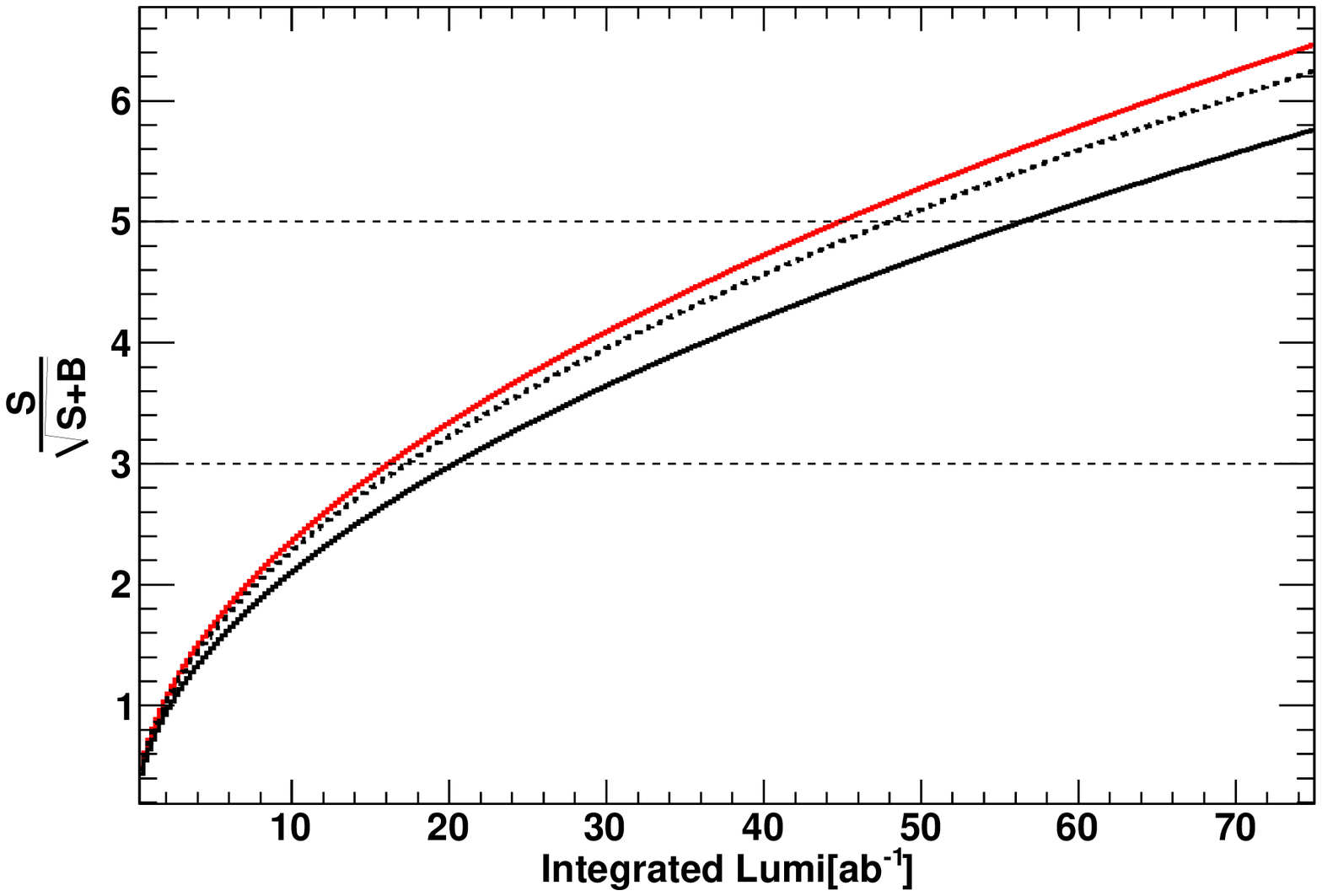}
\includegraphics[width=4.9cm,keepaspectratio]{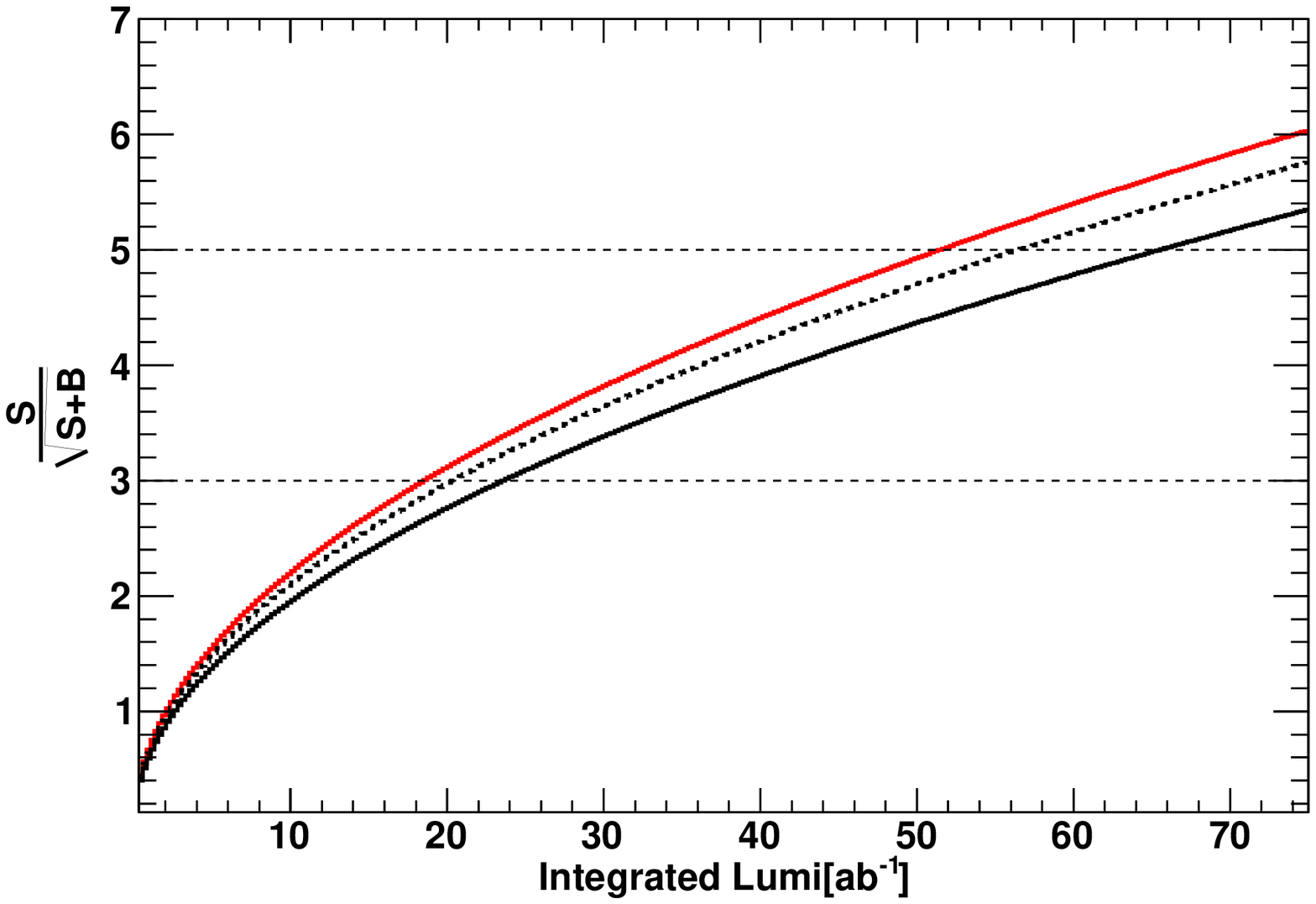}
\includegraphics[width=4.9cm,keepaspectratio]{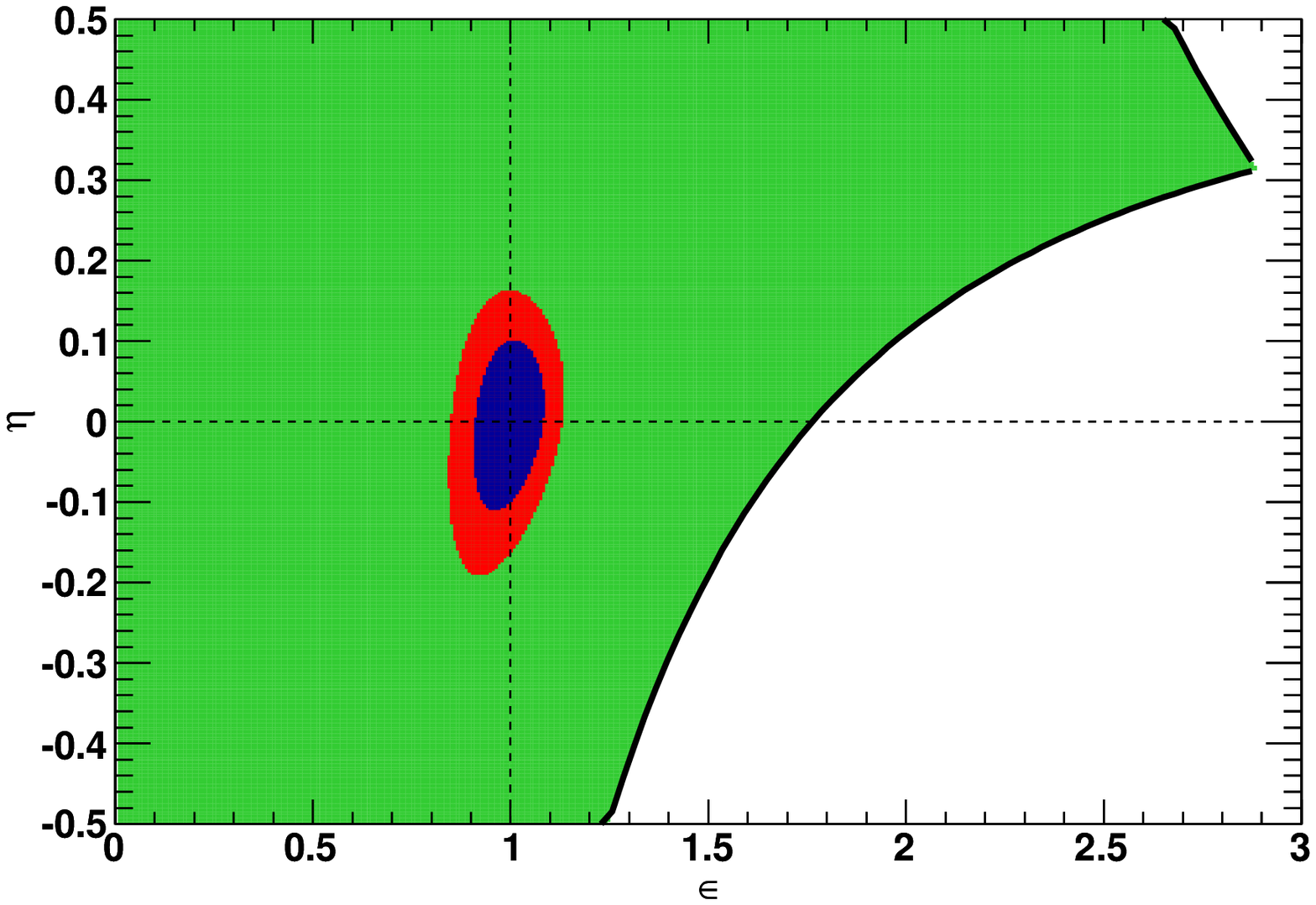}
\end{center}
\caption
{\label{fig:Ext_BFsKnunu_Kstnunu}
{\em Expected sensitivities for the ${\rm Br}(B\rightarrow K\nu\bar{\nu})$ (left) and ${\rm Br}(B\rightarrow K^*\nu\bar{\nu})$ (middle) 
as a function of the integrated luminosity; and expected constraint on the $(\epsilon,\eta)$ plane for the measurement of the before 
mentioned branching ratios at $75{\rm ab^{-1}}$ (right).
}}
\end{figure}

The SuperB fast simulation has been used to produce signal samples in the before mentionned detector configurations, BaBar, 
SuperB base-line and SuperB+Fwd-PID. This test showed a $15\%$ to $20\%$ increase in efficiency using the SuperB+Fwd-PID configuration
with respect to the BaBar setup, depending on the final state, mainly due to the boost reduction. For the time being no generic $B\bar{B}$ 
samples has being produced. To be conservative it has been assumed that the background efficiency increases by the same 
factor as the signal in such a way that the signal to background ratio ($S/B$) stays constant. This global increase in 
efficiency provides a gain on $S/\sqrt{(S+B)}$, which would be interpreted as the signal significance for a cut and count 
analysis. The $S/\sqrt{(S+B)}$ ratio, for both $B\rightarrow K\nu\bar{\nu}$ and $B\rightarrow K^*\nu\bar{\nu}$ modes, as a 
function of the integrated luminosity for the three detector configurations is shown in the left and middle plots of 
figure~\ref{fig:Ext_BFsKnunu_Kstnunu} (BaBar (solid-black), SuperB (dotted-black) and SuperB+Fwd-PID (solid-red)). 
A sensitivity of $15\%$ ($17\%$) are expected for $B\rightarrow K\nu\bar{\nu}$ ($B\rightarrow K^*\nu\bar{\nu}$) at $75{\rm ab^{-1}}$.
The rightmost plot of figure~\ref{fig:Ext_BFsKnunu_Kstnunu} shows the constraint at $68\%$ (blue) and $95\%$ (red) in the 
$(\epsilon,\eta)$ plane for the expected sensitivities on ${\rm Br}(B\rightarrow K\nu\bar{\nu})$ and 
${\rm Br}(B\rightarrow K^*\nu\bar{\nu})$ at $75{\rm ab^{-1}}$.

In summary, it has been investigated the reach of SuperB in the search of the $B\rightarrow K^{(*)}\nu\bar{\nu}$ 
decays with both the hadronic and semi-leptonic techniques. Preliminary results based on the SuperB fast simulation have shown 
an $15$ to $25\%$ increase in the global efficiency with respect to Babar. It has also been shown that SuperB will allow 
an unprecedent reduction of the NP parameter space, $(\epsilon,\eta)$ plane, given the expected sensitivities at $75{\rm ab^{-1}}$ 
of data.

\end{document}